\documentclass[twocolumn,english,aps,prl,superscriptaddress]{revtex4}
\usepackage{amsmath}
\usepackage{graphicx}
\usepackage{snapshot}
\makeatletter

\usepackage{amsfonts}
\usepackage{amsthm}
\usepackage{mathrsfs}
\usepackage{graphics}
\usepackage{lscape}
\usepackage{changebar}
\usepackage{epsfig}
\usepackage{babel}
\usepackage{array}
\usepackage{color}

\makeatother

\begin{document}

\title{Continuum modeling of mechanically-induced creep in dense granular materials}

\author{David L. Henann}
 \email{david_henann@brown.edu}
 \affiliation{School of Engineering, Brown University, Providence, RI 02906, USA}
 \author{Ken Kamrin}
 \email{kkamrin@mit.edu}
 \affiliation{Department of Mechanical Engineering, MIT, Cambridge, MA 02139, USA}

\date{\today}

\begin{abstract}
Recently, a new nonlocal granular rheology was successfully used to predict steady granular flows, including grain-size-dependent shear features, in a wide variety of flow configurations, including all variations of the split-bottom cell. A related problem in granular flow is that of mechanically-induced creep, in which shear deformation in one region of a granular medium fluidizes its entirety, including regions far from the sheared zone, effectively erasing the yield condition everywhere. This enables creep deformation when a force is applied in the nominally quiescent region through an intruder such as a cylindrical or spherical probe. We show that the nonlocal fluidity model is capable of capturing this phenomenology. Specifically, we explore creep of a circular intruder in a two-dimensional annular Couette cell and show that the model captures all salient features observed in experiments, including both the rate-independent nature of creep for sufficiently slow driving rates and the faster-than-linear increase in the creep speed with the force applied to the intruder.
\end{abstract}

\pacs{PACS number(s): }

\maketitle

Cooperativity is a hallmark of quasi-static, dense granular deformation. At a microscopic level, deformation in granular materials takes place through the rearrangement of clusters of grains. These rearrangement events lead to long-range fluctuations, or agitations, that influence the behavior of nearby clusters, leading to macroscopic manifestations of cooperativity, which are quite varied. Most familiarly, the length-scales associated with velocity fields in dense granular flows depend crucially upon the grain size \cite{midi2004}, with grain-size dependent shear band widths being observed in many geometries \cite{mueth2000,komatsu2001,fenistein2003,fenistein2004,fenistein2006,cheng2006,siavoshi2006}. Other manifestations of cooperativity include the dependence of volumetric outflow rate on grain size in drainage flows \cite{beverloo1961,choi2005} and the so-called $H_{\rm stop}$-effect, in which thin granular layers require greater tilt to flow down an inclined surface \cite{pouliquen1999,silbert2003}. A more recently-observed example of cooperativity in dense granular flows is that of mechanically-induced creep \cite{nichol2010,reddy2011,wandersman2014}, understood as follows. When an intruder, such as a sphere, rod, or vane, is placed in a dense granular material, one must apply a force to the intruder that exceeds a critical value in order to move it through the granular media. However, shear deformation far from the intruder enables it to move, or creep, through the granular media for any non-zero value of the applied force -- even when this force is less than the critical value. That is to say, flow anywhere in a granular media erases the yield condition everywhere!

These cooperative phenomena provide stringent tests for a continuum model of granular flow. 
Local approaches to continuum modeling of granular materials, such as granular rheology \cite{dacruz2005,jop2006,kamrin2010} or soil mechanics \cite{schofield1968,nedderman1992}, which relate the stress at a point to the local strain, strain-rate, or locally evolved state variables, are not equipped to address size-dependent manifestations of cooperativity. Recently, we proposed a nonlocal rheology for dense granular flows, capable of modeling size-effects \cite{kamrin2012,henann2013,henann2014}, based on work in the emulsions community \cite{goyon2008,bocquet2009}, which we briefly summarize here. We denote the symmetric strain-rate tensor as $\dot\gamma_{ij} = (1/2)(\partial v_i/\partial x_j + \partial v_j/\partial x_i)$ with $v_i$ the velocity vector and $x_i$ the spatial coordinate. We then assume that steady flow proceeds at constant volume so that $\dot\gamma_{kk} = 0$ \cite{jop2006,rycroft2009,koval2009,kamrin2010} and define the equivalent shear strain rate as $\dot\gamma = (2\dot\gamma_{ij}\dot\gamma_{ij})^{1/2}$. Next, we introduce the symmetric Cauchy stress  $\sigma_{ij}$ and define the pressure $P=-(1/3)\sigma_{kk}$, the stress deviator $\sigma_{ij}' = \sigma_{ij} + P\delta_{ij}$, the equivalent shear stress $\tau=(\sigma_{ij}'\sigma_{ij}'/2)^{1/2}$, and the stress ratio $\mu=\tau/P$. Central to the model is a scalar state variable $g$, called the granular fluidity, which represents the susceptibility of a granular cluster to flow. Mathematically, it functions as a field variable that relates the load intensity $\mu$ to the consequent flow rate $\dot{\gamma}$, i.e., $\dot{\gamma}=g\mu$, so that the tensorial relation between the Cauchy stress and the strain rate is
\begin{equation}\label{eq1}
\sigma_{ij} = -P\delta_{ij} + 2({P}/{g})\dot\gamma_{ij},
\end{equation}
where we have made the common assumption that the strain-rate and deviatoric Cauchy stress tensors are codirectional \cite{jop2006,rycroft2009,kamrin2010}, though this is an approximation \cite{luding2007,depken2007}. In a local description of granular flow, the fluidity is constitutively given as a function of the stress, in a manner consistent with Bagnold scaling \cite{bagnold1954}. Simple dimensional analysis applied to the case of homogeneous simple shearing produces a one-to-one relationship between the stress ratio $\mu$ and a dimensionless version of the strain rate called the inertial number $I=\dot\gamma\sqrt{d^2\rho_{\rm s}/P}$, where $d$ is the mean grain diameter and $\rho_{\rm s}$ is the grain material density. Data has verified a Bingham-like functional form for $\mu=\mu(I)$ \cite{dacruz2005}, which when adopted leads to the following local description of the fluidity:
\begin{equation}\label{eq2}
g_{\rm loc} = {\dot\gamma_{\rm loc}}/{\mu} = H(\mu-\mu_{\rm s})\sqrt{P/\rho_{\rm s}d^2} \ \left[(\mu-\mu_{\rm s})/\mu b\right],  \end{equation}
where $\mu_{\rm s}$ is a static yield value, $b$ is a dimensionless constant characterizing the rate-dependent response of the granular media, and $H$ is the Heaviside step function. 

While the local relation successfully describes homogeneous simple shear, in inhomogeneous flows, significant deviation from this local description is observed \cite{koval2009,kamrin2012}, which motivates a nonlocal differential relation for the granular fluidity. We have proposed the following specific functional form:
\begin{equation}\label{eq3}
\nabla^2 g = ({1}/{\xi^2})\left(g - g_{\rm loc}\right),
\end{equation}
where $\nabla^2(\cdot)$ denotes the Laplacian operator and $\xi(\mu) = Ad/\sqrt{|\mu-\mu_{\rm s}|}$ is the stress-dependent cooperativity length and $A$ is a dimensionless material parameter called the nonlocal amplitude. In the absence of the $\nabla^2 g$ term, the model reduces to the local description with $g = g_{\rm loc}$; however, when flow inhomogeneity is present, this term accounts for nonlocal, cooperative effects and quantitatively captures the loss of local constitutive uniqueness between $\mu$ and $I$.
This nonlocal model was demonstrated to accurately describe flow fields and shear-band widths in numerous granular flow experiments in several distinct flow configurations \cite{henann2013}. While also a manifestation of cooperativity in granular materials, the problem of mechanically-induced creep is distinctly different from the previously-addressed problems, as it probes the nonlocality of strength rather than flow and contains a crucial non-kinematic boundary condition, i.e., the intruder force. As such, it is the purpose of this letter to demonstrate that our model is capable of capturing this phenomenology.

Our present study is motivated by the recent rod-creep experiments of Reddy et al \cite{reddy2011}. In their work, a cylindrical rod was placed vertically in an annular Couette cell filled with grains. When the inner wall was fixed, they observed that the force applied to the rod had to exceed a critical value $F_{\rm c}$ in order to move it through the granular medium.  However, when the inner wall was rotated and an inner-wall-located shear band was formed, the yield condition vanished everywhere, enabling the rod to creep even when the applied force was less than $F_{\rm c}$. (The vanishing of the yield condition was also seen in the experiments of \cite{nichol2010} and \cite{wandersman2014}.) Importantly, several experimental observations were made about the creep phenomenology: (1) For sufficiently low inner wall speeds, mechanically-induced creep is a rate-independent process, i.e., the rod creep speed is linearly proportional to the inner wall speed. (2) The rod creep speed increases exponentially with the force applied to the intruder. (3) The rod creeps faster when it is placed closer to the inner wall shear band.

\begin{figure}
\includegraphics{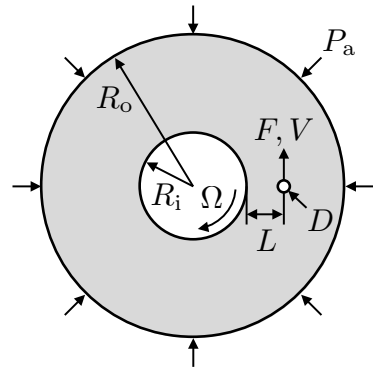}
\caption{Schematic of the two-dimensional analogue of the experiments of \cite{reddy2011}.}\label{fig01}
\end{figure}

In this work, we apply our nonlocal model to the problem of mechanically-induced creep, using finite-element calculations in the commercial package Abaqus/Standard \cite{abaqus}. For computational efficiency, we consider a two-dimensional analogue of the experiments of Reddy et al \cite{reddy2011}, pictured in Fig.\,\ref{fig01}. The geometry is a planar annular shear cell with rough walls at an inner radius $R_{\rm i}$ and outer radius $R_{\rm o}$. The inner wall is specified to rotate at a fixed rate $\Omega$, and the outer wall does not rotate but may move radially so as to impart a confining pressure $P_{\rm a}$. A circular intruder with diameter $D$, which we specify to be rigid and frictionless, is located at a distance $L$ away from the inner wall.  Following \cite{reddy2011}, we take $R_{\rm i}/d=60$, $R_{\rm o}/d = 180$, and $D/d = 2$, throughout, and consider different values of $L/d$. The value of outer radius is sufficiently large so as not to affect the calculation results, consistent with experiments. Finally, in our calculations, either the speed of the intruder $V$ or the force applied to the intruder $F$ is specified. We then calculate the steady flow fields predicted by the nonlocal rheology using Abaqus. The governing partial differential equations are the equilibrium equations $\partial \sigma_{ij}/\partial x_j = 0_i$, where inertia is neglected since we are considering a quasi-static process and there is no gravitational body force since we are in two-dimensions, and the differential relation for the granular fluidity \eqref{eq3}. These are solved in conjunction with the constitutive equations \eqref{eq1} and \eqref{eq2} by means of a user element subroutine (UEL) in Abaqus. The mechanical boundary conditions are as described above, and for the fluidity boundary conditions, we specify that $n_i(\partial g/\partial x_i)=0$ at the inner and outer walls where $n_i$ is the outward surface normal. A detailed discussion of the intruder boundary conditions is given in the Supplemental Material \cite{supp}. The necessary material parameters are $\{\mu_{\rm s},b,A\}$. Following previous work involving glass beads \cite{jop2005,kamrin2010,henann2013}, we take $\mu_{\rm s}=0.3819$ and $b=0.9377$. For the two-dimensional problem, we take $A=1.8$ (see Supplemental Material \cite{supp} for a justification of this selection).

In our first set of simulations, we determine the critical force $F_{\rm c}$. To this end, we fix the inner wall ($\Omega=0$) and specify normalized intruder velocities $V\sqrt{\rho_{\rm s}/P_{\rm a}}$ spanning from $10^{-7}$ to $10^{-2}$. Each calculation quickly reaches a steady-state, and we plot the normalized steady-state applied force  $F/P_{\rm a}D$ as a function of velocity in Fig.\,\ref{fig02} (see Supplemental Material \cite{supp} for a demonstration of mesh insensitivity). This result was observed to be identical for $L/d=18$, 24, and 34. For normalized velocities greater than approximately $10^{-5}$, we see a clear rate dependence. In the calculations, this corresponds to a non-negligible portion of the domain achieving $\mu>\mu_{\rm s}$ so that the $g_{\rm loc}$ term in \eqref{eq3} has a substantial effect in this region. See the contour plot inset of the stress ratio $\mu$ in the region of the intruder in Fig.\,\ref{fig02} corresponding to $V\sqrt{\rho_{\rm s}/P_{\rm a}}=1\times 10^{-3}$. The white region represents the intruder, which is moving upwards, and the region in which $\mu>\mu_{\rm s}$ is denoted as black. Much of the region in front of and behind the intruder has met this condition. However, for $V\sqrt{\rho_{\rm s}/P_{\rm a}}$ less than $10^{-5}$ the force becomes nominally rate-independent, plateauing to a constant value of $F_{\rm c}/P_{\rm a}D = 2.91$, which we denote as the critical force. In this case, the region around the intruder where $\mu>\mu_{\rm s}$ is significantly smaller (see the inset of Fig.\,\ref{fig02} for $V\sqrt{\rho_{\rm s}/P_{\rm a}}=1\times 10^{-7}$). The relation of Fig.\,\ref{fig02} is consistent with analogous experimental observations (see Fig.\,1(c) of \cite{reddy2011} and Fig.\,2(b) of \cite{wandersman2014}).

\begin{figure}
\includegraphics{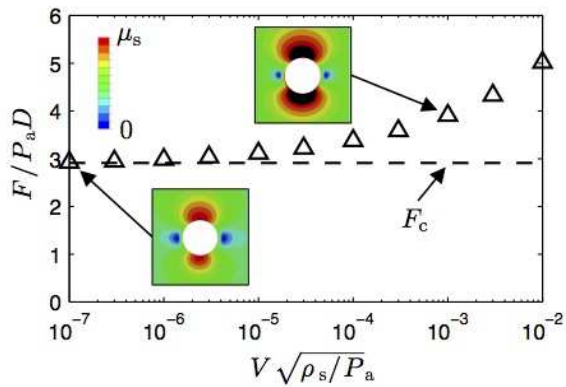}
\caption{Force applied to the intruder as a function of intruder velocity when the inner wall is fixed $\Omega=0$. The dashed line indicates the rate-independent critical force $F_{\rm c}/P_{\rm a}D=2.91$. Insets show contour plots of the stress ratio $\mu$ in the region of the intruder; black regions denote where $\mu>\mu_{\rm s}$.}\label{fig02}
\end{figure}

\begin{figure}[!t]
\includegraphics{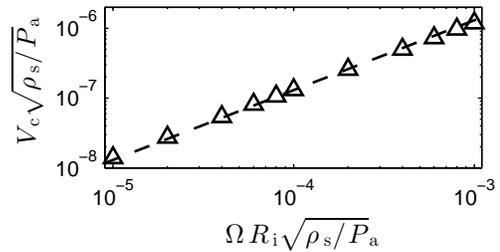}
\caption{Intruder creep speed as a function of inner wall speed for $F/F_{\rm c}=0.75$ and $L/d=24$, demonstrating nominal rate-independence in a range of sufficiently small inner wall speeds; the dashed line denotes the linear relation $V_{\rm c}/\Omega R_{\rm i}=1.3\times 10^{-3}$.}\label{fig03}
\end{figure}

Next, we examine creep phenomenology when the inner wall is rotated. In our calculations, we first apply a force to the intruder less than $F_{\rm c}$ and hold the force constant for a period of time. During this time, the intruder does not creep, and the granular fluidity is zero everywhere. The inner wall is then moved at a fixed speed in a tangential direction which opposes the force applied to the intruder (see Fig.\,\ref{fig01}). This sets up an inner-wall-located shear band and induces a granular fluidity field which is non-zero everywhere. This fluidity field decays with radial distance from the inner wall and is quite small in the region of the intruder. However, fluidization of this region causes the intruder to begin creeping. Steady-state is quickly reached, and we denote the constant creep speed of the intruder as $V_{\rm c}$. Figure \ref{fig03} shows the normalized intruder creep speed $V_{\rm c}\sqrt{\rho_{\rm s}/P_{\rm a}}$ as a function of the normalized inner wall speed $\Omega R_{\rm i}\sqrt{\rho_{\rm s}/P_{\rm a}}$ for a fixed applied force $F/F_{\rm c}=0.75$ and intruder position $L/d=24$, indicating a linear relationship with $V_{\rm c}/\Omega R_{\rm i} = 1.3\times 10^{-3}$. This is a hallmark of rate-independence and demonstrates that the material time-scale $d\sqrt{\rho_{\rm s}/P_{\rm a}}$ is nominally irrelevant for $\Omega R_{\rm i}\sqrt{\rho_{\rm s}/P_{\rm a}}\lesssim 10^{-3}$, and that the time-scale dictating the creep speed of the intruder is that imposed by the inner wall speed. As the inner wall speed is increased further, a deviation from linearity is observed in the calculations, indicating an increased role of the local rheology. This rate-independent regime of creep was observed in all configurations of experiments (see Fig.\,4(a) of \cite{nichol2010}, Fig.\,2(b) of \cite{reddy2011}, and Fig.\,3 of \cite{wandersman2014}).

\begin{figure}[!t]
\includegraphics{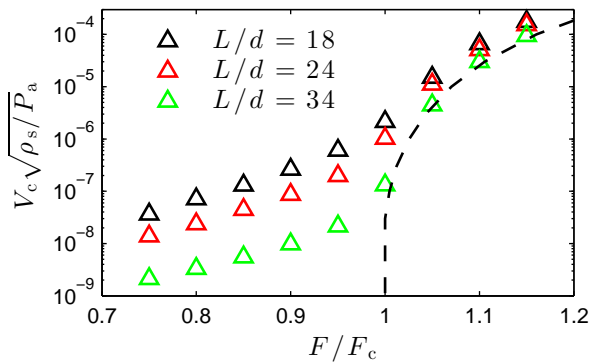}
\caption{Intruder creep speed as a function of applied force for $L/d=18$, 24, and 34 and $\Omega R_{\rm i}\sqrt{\rho_{\rm s}/P_{\rm a}}=1\times 10^{-5}$. Simulated creep data are shown as symbols; the dashed line represents the drag force versus intruder velocity relation of Fig.\,\ref{fig02}.}\label{fig04}
\end{figure}

The creep speed in the rate-independent regime depends on the force applied to the intruder $F/F_{\rm c}$ as well as the distance between the intruder and the inner wall $L/d$. (We did not examine the effect of intruder size $D/d$ in this study.) The calculated dependence is shown as symbols in Fig.\,\ref{fig04} for $L/d=18$, 24, and 34 and $\Omega R_{\rm i}\sqrt{\rho_{\rm s}/P_{\rm a}}=1\times 10^{-5}$. For $F<F_{\rm c}$, we observe that (i) the intruder creeps faster for higher applied force and (ii) the intruder creeps faster when it is closer to the inner wall for all applied forces -- both intuitive notions. What is non-intuitive is that the relation between the creep speed and applied force for $F<F_{\rm c}$ and fixed $L/d$ is nonlinear. In fact, the relation in this range is exponential in character, consistent with experiments (see Fig.\,4 of \cite{reddy2011} and Fig.\,4 of \cite{wandersman2014}). Also plotted in Fig.\,\ref{fig04} as a dashed line is the drag force versus intruder velocity relationship of Fig.\,\ref{fig02}. As the applied force is increased past $F_{\rm c}$, the simulated data for different $L/d$ converge and asymptotically approach the rate-dependent part of this relation, which is dominated by the local rheology rather than nonlocal effects. The convergence of the data around $F=F_{\rm c}$ is qualitatively consistent with experiments \cite{reddy2011}. (For more discussion of this point and additional simulated creep data, see the Supplemental Material \cite{supp}.)

\begin{figure}[!t]
\includegraphics{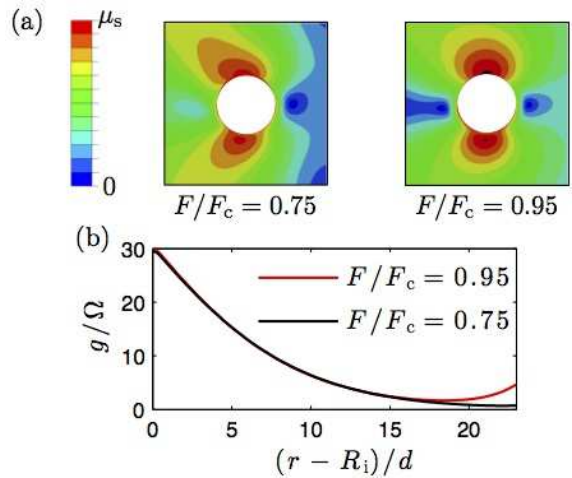}
\caption{(a) Contour plots of the stress ratio $\mu$ in the region of the intruder during steady creep for $F/F_{\rm c}=0.75$ and 0.95, $L/d=24$, and $\Omega R_{\rm i}\sqrt{\rho_{\rm s}/P_{\rm a}}=1\times 10^{-5}$. (b) The granular fluidity along a radial path between the inner wall and the intruder for these two cases.}\label{fig05}
\end{figure}

The faster-than-linear character of the results of Fig.\,\ref{fig04} may be understood using the following examination of the differential relation \eqref{eq3}, while the specifically exponential shape observed is assessed in the Supplemental Material \cite{supp}. The PDE \eqref{eq3} has the structure of an inhomogeneous Helmholtz equation with $g_{\rm loc}$ playing the role of a ``source'' term. When subjected to the boundary condition $n_i(\partial g/\partial x_i)=0$ at the walls, the fluidity field will be zero everywhere if $g_{\rm loc}=0$ (or $\mu<\mu_{\rm s}$) everywhere and will be non-zero everywhere if $g_{\rm loc}>0$ (or $\mu>\mu_{\rm s}$) \emph{anywhere}. In the present simulations, rotation of the inner wall sets up a small region at the inner wall where $g_{\rm loc} >0$ -- a shear band -- and as one moves away from the shear band the fluidity undergoes an exponential-like decay but remains non-zero. This fluidity field may equivalently be thought of as a radially-dependent viscosity field with the viscosity lower closer to the shear band. This justifies the increase of creep speed with decreasing $L/d$. A simplistic Stokes-flow view of the creep problem would indicate that the creep speed and applied force should be linearly related. The fact that this is not observed indicates that the presence of the intruder has an important influence on the fluidity field. This effect is summarized in Fig.\,\ref{fig05}. Figure \ref{fig05}(a) shows contour plots of the stress ratio field $\mu$ in the region of the intruder (which is white and moving upwards) for two values of the applied force $F/F_{\rm c}$, 0.75 and 0.95, and $L/d=24$ and $\Omega R_{\rm i}\sqrt{\rho_{\rm s}/P_{\rm a}}=1\times 10^{-5}$. In both cases, the stress ratio $\mu$ reaches and just exceeds the critical value $\mu_{\rm s}$, leading to $g_{\rm loc}>0$ in a very small region of the intruder boundary. In a sense, this ``activates'' an additional source at the intruder. As can be seen in Fig.\,\ref{fig05}(a), while this region is indiscernible for $F/F_{\rm c}=0.75$, more of the region surrounding the intruder reaches $\mu_{\rm s}$ for the case of $F/F_{\rm c}=0.95$, and as a consequence, the ``source'' is stronger and the fluidity is increased in the region of the intruder. Figure \ref{fig05}(b) shows the fluidity $g$ along a radial path from the inner wall to the intruder for $F/F_{\rm c}=0.75$ and 0.95. Near the inner wall, away from the intruder, the fluidity field is essentially the same in both cases. However, in the region of the intruder, the stronger source in the higher force case leads to a local fluidity which is approximately six times greater -- or equivalently a viscosity which is six times less. This decrease in viscosity combined with the greater applied force leads to the faster than linear increase in creep speed with force observed in Fig.\,\ref{fig04}.

In conclusion, we have demonstrated that the nonlocal granular rheology is capable of describing the phenomenology of mechanically-induced creep observed in experiments \cite{nichol2010,reddy2011,wandersman2014}. The success of the nonlocal granular rheology in capturing these behaviors lends confidence in the model. It bears noting that the model was not developed with creep phenomenology in mind. The fact that the model passes both the test of describing shear band widths in dense granular flows \cite{henann2013} as well as mechanically-induced creep demonstrates the generality of the model.

DLH acknowledges funds from the Brown University School of Engineering, and KK acknowledges funds from NSF-CBET-1253228 and the MIT Department of Mechanical Engineering.

\section{Supplemental material}

{\it Intruder boundary conditions}:
Here, we give a detailed description of the boundary conditions at the intruder. To model frictionless interaction between grains and the rigid circular intruder, we constrain a ring of nodes representing the intruder boundary to move along but not through the circular boundary. This boundary condition is achieved by first creating a reference point for the intruder at its center and then specifying that the distance between the reference point and nodes on the circular intruder boundary remain fixed through use of a link-type multi-point-constraint in Abaqus \cite{abaqus}. Then, either the intruder speed $V$ or the force $F$ is applied directly to the reference point associated with the intruder. While the intruder boundary is specified to be mechanically rigid and frictionless, the domain in which we solve the governing equations includes the intruder, so as to circumvent the need for an extra fluidity boundary condition. We have also performed additional calculations using a $n_i(\partial g/\partial x_i)=0$ fluidity boundary condition at the intruder boundary as well as calculations with a mechanically rough intruder (results not shown). We observe no qualitative change in the simulated results in either case, indicating that the results reported in the letter are robust. That is to say, the model is capable of capturing the phenomenology of mechanically-induced creep regardless of the details of the specified intruder boundary conditions.

{\it Nonlocal amplitude $A$ for a two-dimensional setting}:
The nonlocal amplitude $A$ is a crucial material parameter in the nonlocal rheology, setting the widths of shear bands in the rate-independent limit. In our past work \cite{henann2013}, we found that a value of $A=0.48$ provided an excellent description of three-dimensional flows of glass beads; however, we expect that a different quantitative value will be appropriate for our effective two-dimensional calculations. Since we are using the rod creep experiments of Reddy et al.\,\cite{reddy2011} as a guide in our calculations (for example, in our choice of geometry), we use their annular shear experiments (see their Fig.\,1(b)) to guide our selection of $A$. 

\begin{figure}[!b]
\includegraphics{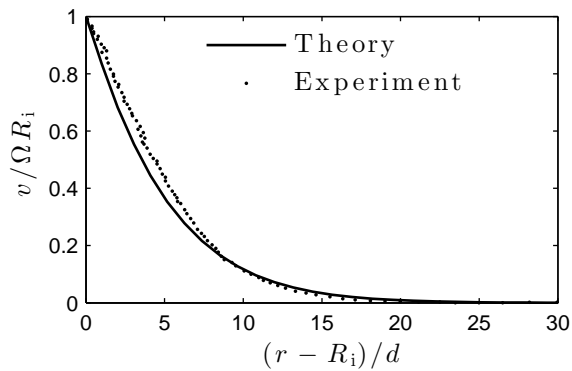}
\caption{Comparison of calculation results in two dimensional annular shear to the three-dimensional annular shear experiments of Reddy et al.\,\cite{reddy2011}. This comparison was used as a guide to select $A=1.8$.}\label{figS01}
\end{figure}

We perform simulations of two-dimensional annular shear with \emph{no intruder}. The annular cell has rough walls at inner radius $R_{\rm i}$ and outer radius $R_{\rm o}$. As in all the calculations in this letter, the inner wall is specified to rotate at a fixed rate $\Omega$, which gives rise to a wall-located shear band, and the outer wall does not rotate but may move radially so as to impart a confining pressure $P_{\rm a}$. Following Reddy et al.\,\cite{reddy2011}, we take $R_{\rm i}/d=60$ and $R_{\rm o}/d = 180$; however, the value of $R_{\rm o}$ does not affect calculation results. We find that a normalized inner wall speed of $\Omega R_{\rm i}\sqrt{\rho_{\rm s}/P_{\rm a}}=1\times 10^{-5}$ is sufficiently slow so as to yield a rate-independent result. The calculated flow field is tangentially symmetric, and the steady-state tangential velocity field $v$, normalized by the inner wall speed $\Omega R_{\rm i}$, is plotted in Fig.\,\ref{figS01} as a solid line. We have selected $A=1.8$ so that the calculated result matches the experiments of Reddy et al.\,\cite{reddy2011} (plotted as points in Fig.\,\ref{figS01}).

\begin{figure}[!t]
\includegraphics{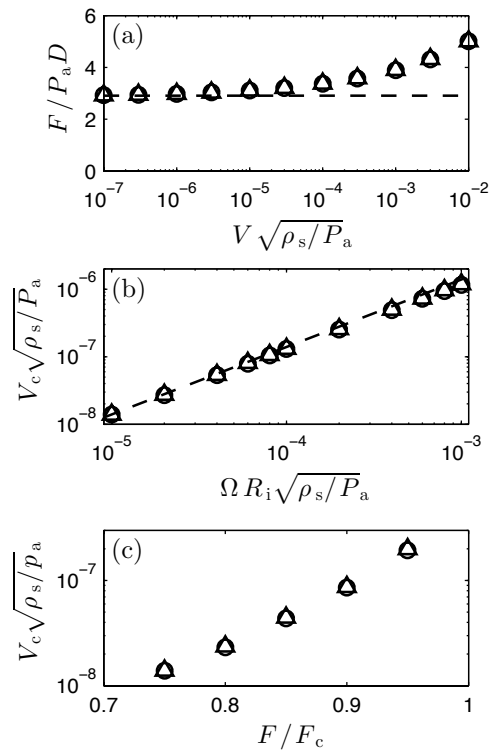}
\caption{Results of the mesh sensitivity study for (a) rod dragging, (b) rod creep for different inner wall speeds and $F/F_{\rm c}=0.75$, and (c) rod creep for different applied forces and $\Omega R_{\rm i}\sqrt{\rho_{\rm s}/P_{\rm a}}=1\times 10^{-5}$. In all cases, $L/d=24$. We denote fine mesh calculations with a $\triangle$ and coarse mesh calculations with a $\circ$.}\label{figS02}
\end{figure}

\begin{figure}[!t]
\includegraphics{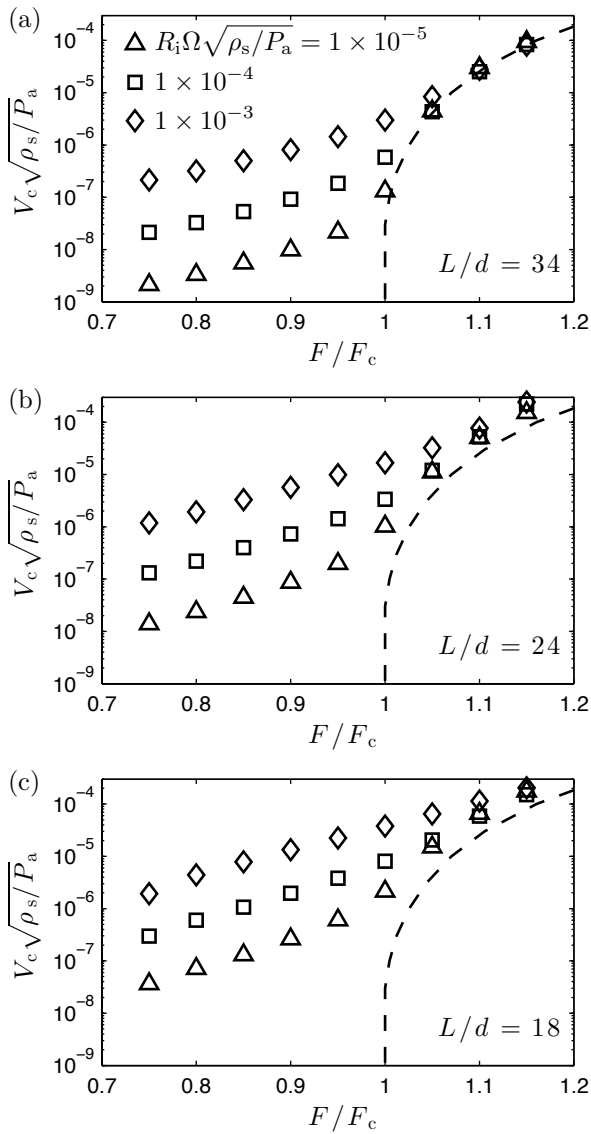}
\caption{Intruder creep speed as a function of applied force for $\Omega R_{\rm i}\sqrt{\rho_{\rm s}/P_{\rm a}}=1\times 10^{-5}$, $1\times 10^{-4}$, and $1\times 10^{-3}$ and (a) $L/d=34$, (b) 24, and (c) 18. Simulated creep data are shown as symbols, and the dashed line represents the drag force versus intruder velocity relation when the inner wall is fixed ($\Omega=0$).}\label{figS03}
\end{figure}

{\it Mesh sensitivity study}:
It is important to verify that our finite-element calculation results do not depend on the mesh resolution. To this end, we reconsider all calculation results for the $L/d=24$ geometry with two mesh resolutions. For the ``coarse'' mesh, we take a mesh resolution of $0.23d$ at the intruder boundary with a total of 77121 elements, and the ``fine'' mesh has a resolution of $0.17d$ at the the intruder boundary and a total of 106212 elements. We first repeat the rod dragging calculations for both meshes, plotted in Fig.\,\ref{figS02}(a). Fine mesh results are plotted with a $\triangle$ and coarse mesh results with a $\circ$. The results are indistinguishable, and both yield a rate-independent plateau of $F_{\rm c} = 2.91 P_{\rm a}D$. The creep speed as a function of normalized inner wall speed for $F/F_{\rm c}=0.75$ is shown in Fig.\,\ref{figS02}(b), and again, the results for the two meshes are indistinguishable, both demonstrating nominal rate-independence indicated by the dashed line. Finally, the intruder creep speed as a function of applied force for $\Omega R_{\rm i}\sqrt{\rho_{\rm s}/P_{\rm a}}=1\times 10^{-5}$ is shown in Fig.\,\ref{figS02}(c), also demonstrating mesh insensitivity.

{\it Additional simulated creep data}: 
In the letter, we reported rod creep calculations for $L/d=18$, 24, and 34 and $\Omega R_{\rm i}\sqrt{\rho_{\rm s}/P_{\rm a}}=1\times 10^{-5}$. Here we show the full results of our rod creep simulations in Fig.\,\ref{figS03}. Figure \ref{figS03}(a) shows the intruder creep speed as a function of the applied force in the range $0.75\le F/F_{\rm c} \le 1.15$ for three different inner wall speeds and $L/d=34$. Figures \ref{figS03}(b) and (c) show the same results for $L/d=24$ and 18. Rate-independent behavior is evident for $F/F_{\rm c}\le 0.95$ and for all inner wall speeds and intruder locations considered here. When the inner wall speed is increased by a decade, the intruder creep speed correspondingly increases by a decade in this range. For $F/F_{\rm c}\ge 1$, all data begin to converge and asymptotically approach the drag force versus intruder velocity relation when the inner wall is fixed (dashed line). This transition indicates an increased role of the material time scale $d\sqrt{\rho_{\rm s}/P_{\rm a}}$ and a dominance of the local rheology in the observed response.

It bears noting that the convergence of the data around $F/F_{\rm c}=1$ is consistent with the experimental observations of Reddy et al.\,\cite{reddy2011}. However, our calculations are quantitatively different. While the data for different $L/d$ converge precisely at $F/F_{\rm c}=1$ in the experiments, the transition is more diffuse in our calculations. We speculate that the quantitative differences in the transition are due to the two-dimensional setting of the simulations compared to the three-dimensional nature of the experiments.

{\it Potential explanations for exponential-type behavior relating force and creep-rate}: 
We have arrived at two preliminary explanations for why the relationship between force and creep-rate bears and exponential-type shape when $F<F_{\rm c}$.  These are not mutually exclusive -- both could be taking place together -- but they remain somewhat speculative at this stage.   The difficulty in constructing such an argument is related to the fact that although this geometry is simple to the eye, it develops complex two-dimensional fields, and reducing these in a way that can be understood with simple mathematical arguments is actually quite non-trivial.  While it is certainly true that the local fluidity source that develops ahead of the intruder gives rise to non-linear behavior,  which we discuss in the main body of the text, the arguments below could shed light on why an exponential-type appearance arises in the process.

A tool in developing these explanations is the integral representation of fluidity solutions. Given a stress field, fluidity solutions can be expressed through the body-integral
$$
g(\mathbf{x})=\int g_{\rm loc}\left(\mathbf{x}-\mathbf{x}'\right)\ G(\mathbf{x}')\ dV'
$$
where $G(\mathbf{x})=G[\xi(\mu(\mathbf{x}))]$ is a Green function for Eq.\,3 in the main text, arising as a global functional operator on $\xi(\mu(\mathbf{x}))$. The Helmoltz-type nature of the fluidity PDE causes the Green function to decay in an exponential fashion, which causes fluidity solutions to exponentially decay away from sources of local fluidity.  For instructive purposes, note that in one-dimension of space and for $\mu={\rm const}$, the infinite-space Green function obeys $G(x)\propto \exp(-|x/\xi(\mu)|)$.

The first explanation for the exponential dependence of the intruder creep relates to the stress-dependence of $\xi$, and how increasing the intruder-force slows the rate of fluidity decay between the objects.   As noted above, the decay length of the Green function is not a constant, but given by the stress through $\xi(\mu)$, with $\xi$ increasing with $\mu$ (when $\mu<\mu_{\rm s}$).  Increasing the force on the rod directly increases the $\mu$ field between the rotating wall and the intruder.  In the large zone of $g_{\rm loc}=0$ between the objects, this increase in $\mu$ directly increases $\xi$, which slows the rate of exponential decay of the fluidity between the two objects. This logic could then explain why the fluidity at the intruder gains an exponential dependence on the intruder-force.   

The second explanation relates to how forcing on the intruder brings the sources of fluidity about the two objects closer together in space.  As force on the rod increases, it causes $\mu$ to increase in a decaying fashion away from the intruder, as we previously discussed.  In so doing, there will be some growth of the region of local fluidity about the intruder and also about the inner wall.  Zones of local fluidity are a source term in the body-integral above, and due to the exponential behavior of the Green's function, if two sources of fluidity get closer, the fluidity values at the two locations increase exponentially.  Consequently, this effect could be at the heart of the exponential dependence of creep-rate on the intruder-force.

We reiterate that these are still preliminary explanations.  We cannot claim that either effect will produce a perfect exponential, due to intricacies of the stress field that have been handled only qualitatively in the arguments just made.

\end{document}